\documentclass[10pt,oneside,reqno]{amsart}
\usepackage{amsfonts,amssymb, amscd,amsmath,latexsym,amsbsy,bm}
\numberwithin{equation}{section}
\usepackage{mathrsfs}
\usepackage{dsfont}
\usepackage{braket}
 \usepackage[usenames,dvipsnames]{xcolor}
 \usepackage[titletoc]{appendix}

\usepackage[colorlinks,hyperindex, % pagebackref,
           bookmarks=true,bookmarksopen=true]{hyperref}
           \hypersetup    {
        colorlinks=true,
       linkcolor=black,
        urlcolor=MidnightBlue,
        citecolor=RedOrange,
   }

\usepackage{epic}
\usepackage{eepic}
\usepackage{graphicx}

\usepackage{pgf}%,pgfarrows}
\usepackage{tikz}

\usetikzlibrary{shapes}
\usetikzlibrary{plotmarks}

\usepackage{cite}
\usepackage{enumerate}
\usepackage[curve]{xy}
\usepackage{stmaryrd}

\usepackage{mathtools}

\renewcommand{\author}[1]{\large\rm #1\\ \bigskip}
\renewcommand{\title}[1]{\bigskip\bigskip\Large\bf #1\bigskip\bigskip\\}

\newcommand{\Bigpsi}[3]{\phantom{\Psi}_2 \kern -.05em
\Psi_2\left(\genfrac{}{}{0pt}{}{#1}{#2}\biggl|#3\right)}

\newcommand{\bea}{\begin{eqnarray}}
\newcommand{\eea}{\end{eqnarray}}

\newcommand{\beq}{\begin{equation}}
\newcommand{\eeq}{\end{equation}}

\renewcommand{\textcolor}[1]{}

 {\begin{list}{}%
         {\setlength{\leftmargin}{#1}}%
         \item[]%
 }
 {\end{list}}

\usepackage{mathtext}
\usepackage{stackrel}

\usepackage[utf8]{inputenc}

\begin{document}

$\;$\\
\vspace{2.7cm}

\begin{center}
{\LARGE \bf Three-body decays of $\pi_{2}(1670)$ and $\rho_{3}(1690)$ via the Sill distribution }
\end{center}

\vspace{0.2cm}

\vskip 1.3 cm

\centerline{\large {\bf Shahriyar Jafarzade$^{1,2}$ and Enrico Trotti$^{1}$}  }

{\small
\begin{center}
\textit{ $^{1}$Institute of Physics, Jan Kochanowski University,\\ ul. Uniwersytecka 7, 25-406, Kielce, Poland} , \\
\vspace{2.5mm} 
\textit{ $^{2}$ Institute of Radiation Problems, Ministry of Science and Education,\\
B.Vahabzade St. 9, AZ1143, Baku, Azerbaijan} \\
\texttt{} \\
\vspace{.1mm}
\vspace{.1mm}
\end{center}
}

\vskip 0.5cm \centerline{\bf Abstract} \vskip 0.2cm \noindent 

 We present the three-body decays for the resonances $\pi_{2}(1670)$ and $ \rho_3(1690)$. We use an effective model based on the flavor symmetry and have assumed the \text{Sill} distribution for describing the lie-shape of intermediate unstable resonances. The vector-pseudoscalar meson decay channel is considered as an intermediate step to explain the decay of three-pseudoscalar meson channels.

\section{Introduction}
\label{intro}
The light ground state mesons, such as $\pi_{2}(1670)$ and $ \rho_3(1690)$ (with $J^{PC}=2^{-+}\,,\, 3^{--}$ ), according to the Particle Data Group (PDG) \cite{Workman:2022ynf}, are well-established. Their total decay widths are $\Gamma^{\text{tot}}_{\pi_2(1670)}=258^{+8}_{-9}$ MeV and $\Gamma^{\text{tot}}_{\rho_3(1690)}=161\pm 10$ MeV.  Experimental data for the decay processes of these resonances was used in \cite{Koenigstein:2016tjw,Giacosa:2023fdz,Jafarzade:2021vhh,Shastry:2021asu} to test the corresponding spontaneous and anomalous breaking features of QCD symmetries used in the extended Linear Sigma Model (eLSM) \cite{Parganlija:2012fy}. We list some of the vector-pseudoscalar meson decay data of these resonances in Tab. \ref{tab:rho3}.  %The amount of experimental data available \textbf{is enough to } make the test of effective models on this resonance phenomenologically interesting. \textbf{Among the effective models constructed, we remind the one presented in Ref. \cite{Jafarzade:2021vhh}, whose provided results are partially reported in Tab. \ref{tab:rho3}}.
\begin{table}[h] %[ptb]
		\centering
		\renewcommand{\arraystretch}{1.5}
		\begin{tabular}[c]{|c|c|c|}
			\hline
			Decay process & Model   \cite{Koenigstein:2016tjw,Jafarzade:2021vhh} & PDG \cite{Workman:2022ynf}   %& LQCD   \cite{Johnson:2020ilc} 
   \\
			\hline 
   $\,\;\pi_2(1670) \rightarrow \rho(770)\, \pi$ &  $80.6\pm10.8$ & $80.6\pm10.8$   %& $-$
   \\\hline
    $\,\;\pi_2(1670) \rightarrow \bar{K}^\ast(892)\, K +\textbf{c}.\mathrm{c}$ &  $11.7\pm 1.6$ & $10.9\pm 3.7$   %& $-$
   \\\hline
    $\,\;\rho_3(1690) \rightarrow \omega(782)\, \pi$ &  $35.8\pm7.4$ & $25.8\pm 9.8$ % & $22$  
   \\
			\hline
				$\,\;\rho_3(1690) \rightarrow \rho(770)\, \eta$ &  $3.8\pm0.8$ & $-$  % & $-$  
    \\
			\hline
  						$\,\;\rho_3(1690) \rightarrow \bar{K}^\ast(892)\, K +\textbf{c}.\mathrm{c}.$ & $ 3.4\pm 0.7$  & $-$ % & $2$
      \\
			\hline
	%$K_{3}^{\ast} (1780) \rightarrow \rho(770) \, K$									&	$16.8 \pm 3.5$					&	$49.3 \pm 15.7$ %& $-$ 
% \\
	%		\hline
	%		$K_{3}^{\ast} (1780) \rightarrow \bar{K}^{\ast} (892) \, \pi$						&	$27.2 \pm 5.6$					&	$31.8 \pm 9.0$ %& $-$
  % \\ \hline
 %  $\omega_{3} (1670) \rightarrow \rho (770) \, \pi$									&	$97 \pm 20$						&	seen & 62 \\
	%		\hline
  % $\omega_{3} (1670) \rightarrow \bar{K}^{\ast} (892) \, K$							&	$2.9 \pm 0.6$ & $-$ &  2 \\ \hline
   %$\phi_{3} (1850) \rightarrow \rho (770) \, \pi$										&	$1.1 \pm 0.2$					& $-$ & $-$
	%			\\ \hline
    %$\phi_{3} (1850) \rightarrow \bar{K}^{\ast} (892) \, K$								&	$35.5 \pm 7.3$					&	seen & 20 \\\hline
		\end{tabular}
		\caption{Decay rates (MeV) of $\pi_2(1670)$ and $\rho_3(1690)$ .  (The first entry \\has been used to determine the model parameter.)}
		\label{tab:rho3}
				\end{table}\\
%The $\rho_3(1690)$ meson has been recently studied with QCD sum rules \cite{Aliev:2023tqy} and Lattice QCD \cite{Johnson:2020ilc}.
In this work, we present novel results for three-body decays of these resonances based on the effective model constructed in Refs. \cite{Koenigstein:2016tjw,Jafarzade:2021vhh}. To this end, we use the \text{Sill} distribution developed in Ref. \cite{Giacosa:2021mbz} and successively used in the case of e.g. spin-2 mesons \cite{Jafarzade:2022uqo} and hybrid mesons in Ref. \cite{Shastry:2023ths} (for other works that have applied the \text{Sill} distribution, see also \cite{Winney:2022tky,Yildirim:2023znd,ALICE:2023wjz,Giacosa:2023xti}).
\section{Effective Model}
\label{sec-1}
The following effective interaction term describes the decay of the pseudotensor mesons\footnote{$P_2$ is a $3 \times 3$ matrix containing the nonet of listed states, see \cite{Koenigstein:2016tjw}. The same is true for the other quoted nonets.} $\Big(P_2$=$\Big\{\pi_2(1670), K_2(1770)$,\\$\eta_2(1870)$, $\eta_2(1645)\Big\}\Big)$  into the vector $\Big(V$=$\Big\{\rho(770), K^{\ast}(892), \omega(782), \phi(1020)\Big\}\Big)$ and pseudoscalar $\Big(P$=$\Big\{\pi,K,
\eta(547),\\\eta^{\prime}(958)\Big\}\Big)$ mesons \cite{Koenigstein:2016tjw}:
\begin{equation}
\mathcal{L}_{p_2vp}=g_{p_2vp}\,\mathrm{tr}\Big[P_2^{\mu\nu}\big[V_{\mu}\,,(\partial_\nu P)\big]_{-}\Big]\,\text{ ,}%
\label{eq:lag-wsp}%
\end{equation}
which leads to the decay rate:
\begin{equation}
\Gamma_{P_2\rightarrow V+P}(m_{p_2},m_{v},m_{p})=\frac
{g_{p_2vp}^2\,|\vec{k}_{p_2,v,p}|^{3}}{120\,\pi\,m_{p_2}^{2}}
\Big(5+\frac{2\,|\vec{k}_{p_2,v,p}|^{2}}{m_{v}^{2}}\Big)\,\kappa
_{i}\,\Theta(m_{p_2}-m_{v}-m_{p})\,.
\end{equation}
Above $\kappa_i$ is a decay coefficient that takes into account flavor combinatoric and renormalization factors, extracted from the extended version of the lagrangian. The coupling constant is determined to be:
\begin{align}
			g_{p_2 v p}^{2} = ( 11.9 \pm 1.6 )  	\,.
		\end{align}
A similar Lagrangian describing the decay of spin-$3$ tensor mesons $\Big(W_3$=$\Big\{\rho_3(1690),K_{3}^{\ast} (1780), \phi_{3} (1850), \omega_{3} (1670)\Big\}\Big)$ has the following form \cite{Jafarzade:2021vhh}:
		\begin{align}\label{eq:Lag-wvp}
			 \mathcal{L}_{w_3 v p} &=	
			 g_{w_3 v p} \, \varepsilon^{\mu\nu\rho\sigma} \, \mathrm{tr} \Big[ W_{3 , \mu\alpha\beta} \, \big\{ ( \partial_\nu V_{\rho} ) , \, ( \partial^{\alpha} \partial^{\beta} \partial_{\sigma} P ) \big\}_{+} \Big] \, .
    %\\\nonumber
   %& =\frac{g_{w_3 v_1 p}}{4} \, \varepsilon^{\mu\nu\rho\sigma} \,\Big(\rho^0_{3 , \mu\alpha\beta} \Big\{ - (\partial_\nu \bar{K}^{\ast 0}_\rho) (\partial^\alpha \partial^\beta \partial_\sigma K^0) 
%			- (\partial_\nu K^{\ast 0}_\rho) (\partial^\alpha \partial^\beta \partial_\sigma \bar{K}^0) 
 %   + (\partial_\nu K^{\ast +}_\rho) (\partial^\alpha \partial^\beta \partial_\sigma K^-) \\	\nonumber
  % & +
   % (\partial_\nu K^{\ast -}_\rho) (\partial^\alpha \partial^\beta \partial_\sigma K^+)  + 2 \, ( \partial_\nu \rho_\rho^0 ) ( \partial^\alpha \partial^\beta \partial_\sigma \eta ) \cos \beta_p\Big\}\Big) +\text{other terms}    
		\end{align}
%where the explicit form of the meson nonet reads
%	\begin{align}\nonumber
%			&P = \frac{1}{\sqrt{2}}
%			\begin{pmatrix}
%				\frac{\eta_{N} + \pi^{0}}{\sqrt{2}}	&	\pi^{+}								&	K^{+}		\\
%				\pi^{-}								&	\frac{\eta_{N} - \pi^{0}}{\sqrt{2}}	&	K^{0}		\\
%				K^{-}								&	\bar{K}^{0}							&	\eta_{S}
%			\end{pmatrix} \, ,\,\,\,
%			 V^{\mu} = \frac{1}{\sqrt{2}}
%			\begin{pmatrix}
%				\frac{\omega_{1 , N} + \rho_1^{0 }}{\sqrt{2}}	&	\rho_1^{+ }										&	K_1^{\ast + }	\\
%				\rho_1^{-}										&	\frac{\omega_{1 , N} - \rho_1^{0 }}{\sqrt{2}}	&	K_1^{\ast 0}	\\
%				K_1^{\ast - }										&	\bar{K}_1^{\ast 0 }								&	\omega_{1 , S}
%			\end{pmatrix}^{\mu} \,,\\ \\
 %  			& \qquad \qquad W_3^{\mu\nu\rho} = \frac{1}{\sqrt{2}}
%			\begin{pmatrix}
%				\frac{\omega_{3 , N} + \rho_{3}^{0 }}{\sqrt{2}}	&	\rho_{3}^{+ }													&	K_{3}^{+ }		\\
%				\rho_{3}^{- }													&	\frac{\omega_{3 , N} - \rho_{3}^{0 }}{\sqrt{2}}	&	K_{3}^{0 }		\\
				%K_{3}^{- }													&	\bar{K}_{3}^{0 }												&	\omega_{3 , S}
			%\end{pmatrix}^{\mu\nu\rho} \, .	\label{eq:tensor3_nonet}
		%\end{align}
The tree-level decay rate is:
\begin{align}
			 \Gamma_{W_3 \rightarrow V + P} ( m_{w_3} , m_{v} , m_{p} ) =	 g_{w_3 v p}^{2} \, \frac{\big| \vec{k}_{w_3,v , p} \big|^{7}}{105} \, \kappa_{i} \, \Theta( m_{w_3} - m_{v} - m_{p} ) \, .
		\end{align}
  As a consequence of the parameter determination shown in Ref. \cite{Jafarzade:2021vhh}, one has the following value for the coupling:
\begin{align}
			g_{w_3 v p}^{2} = ( 9.2 \pm 1.9 ) \cdot 10^{-16} \, \, \text{MeV}^{-6}	\,.
   \label{eq:coupling_w3_v1_p}
		\end{align}
%as well as the decay rates
%\begin{align*}
 %   \Gamma_{\rho_{3} (1690) \rightarrow \rho (770) \, \eta}=3.8 \pm 0.8 \,\text{MeV}\,,\qquad \Gamma_{\rho_{3} (1690) \rightarrow \bar{K}^{\ast} (892) \, K}=3.36 \pm 0.69 \,\text{MeV}\,,
%\end{align*}
Some decay channels are listed in Tab.\ref{tab:rho3} (see Ref. \cite{Jafarzade:2021vhh} for more details).

%In an analogous way, we evaluate the decay channels  \\$\rho_3(1690) \rightarrow \bar{K}^\ast(892)\, K\rightarrow K \overline{K}\pi$ and $\rho_3(1690)\rightarrow \rho\eta\rightarrow \eta\pi\pi$ in the following section.
  
%\section{\text{Sill} Distribution}

\section{Results for three-body decays}

The PDG lists the following decay rates \cite{Workman:2022ynf}:
\begin{align}
  \Gamma_{\pi_2(1670) \rightarrow \pi^{\pm}\pi^{+}\pi^{-}}=137\pm 11\,\text{MeV} \,,\qquad \Gamma_{\rho_3(1690) \rightarrow K \overline{K}\pi}=6.1\pm 2.0\,\text{MeV} \,.
\end{align} 
In order to theoretically describe them, we introduce the ``\text{Sill}'' spectral function $d^{\text{\text{Sill}}}(x)$, which is chosen because of the following features: it is normalized even for broad states, it has a vanishing real part contribution to the loop of virtual particles, it takes into account the decay threshold, and is rather simple to use. For further details, see Ref. \cite{Giacosa:2021mbz,Giacosa:2023xti}. \\
Considering the spectral function for $\rho(770)$, one has:
\begin{align}
    d^{\text{\text{Sill}}}_{\rho}(y)=\frac{2y}{\pi}\frac{\sqrt{y^2-4m_{\pi}^2}\,\tilde{\Gamma}_{\rho}}{(y^2-m_{\rho}^2)^2+(\sqrt{y^2-4m_{\pi}^2}\,\tilde{\Gamma}_{\rho})^2}\,\Theta
(y-2m_{\pi})\,, \hspace{1.2cm} \int_{2m_{\pi}}^{\infty}dy\,d^{\text{Sill}}_{\rho}(y)=1
\text{ ,}
\end{align}
where the definition of $\Tilde{\Gamma}_\rho $ is
\begin{align}
    \tilde{\Gamma}_{\rho}\equiv \frac{\Gamma^{\text{PDG}}_{\rho\rightarrow 2\pi}\,m_{\rho}}{\sqrt{m_{\rho}^2-4m_{\pi}^2}}\;
    \text{ ,}
\end{align}
with the PDG values $\Gamma^{\text{PDG}}_{\rho\rightarrow 2\pi} = 149.1$ MeV and $m_{\rho} =775.11 $ MeV. Then, upon integrating over it, we obtain the following result for the three-body decay of $\pi_2$:
\begin{align}
    \Gamma_{\pi_2\rightarrow \rho\pi\rightarrow \pi\pi\pi}\simeq \int_{2m_\pi}^{\infty}dy\,d^{\text{Sill}}_{\rho}(y) \,\Gamma_{\pi_2\rightarrow \rho\pi}(m_{\pi_2},y,m_{\pi})\simeq (73.9\pm9.9) \,\text{MeV .}
\end{align}
The \text{Sill} distribution for the vector $K^{\star}(892)$-meson reads:
\begin{align}
    d^{\text{Sill}}_{K^\star}(y)&=\frac{2y}{\pi}\frac{\sqrt{y^2-(m_K+m_{\pi})^2}\,\tilde{\Gamma}_{K^\star}}{(y^2-m_{K^\star}^2)^2+(\sqrt{y^2-(m_K+m_{\pi})^2}\,\tilde{\Gamma}_{K^\star})^2}\,\Theta
(y-m_K-m_{\pi})\,,
\end{align}
 where $\Tilde{\Gamma}_{K^\star} $ is linked to  the PDG values $\Gamma^{\text{PDG}}_{K^\star\rightarrow K\pi} = 51.4$ MeV and $m_{K^\star} =890 $ MeV according to:
\begin{align}
    \tilde{\Gamma}_{K^\star}\equiv \frac{\Gamma^{\text{PDG}}_{K^\star\rightarrow K\pi}\,m_{K^\star}}{\sqrt{m_{K^\star}^2-(m_{K}-m_{\pi})^2}}\;
    \text{ .}
\end{align}  
The  normalization 
\begin{align}
\int_{m_k+m_\pi}^{\infty}dy\,d^{\text{Sill}}_{K^\star}(y)=1\,,
\end{align}
holds also here.\\
In this case, the three-body decay of $\rho_{3}(1690)$ can be calculated as:
\begin{align}
    \Gamma_{\rho_3(1690) \rightarrow \bar{K}^\ast(892)\, K\rightarrow K \overline{K}\pi}&\simeq \int_{m_k+m_\pi}^{\infty}dy\,d^{\text{Sill}}_{K^\star}(y) \,\Gamma_{\rho_3\rightarrow \bar{K}^\ast\, K}(m_{\rho_3},y,m_{K})\simeq (3.43\pm 0.70) \,\text{MeV .}
\end{align}
The decay channel $ \Gamma_{\rho_3 \rightarrow\eta\pi\pi}$ has also been seen experimentally \cite{Fukui:1988mp}; we estimate it as:
\begin{align}
    \Gamma_{\rho_3(1690)\rightarrow \rho(770)\eta\rightarrow \eta\pi\pi}\simeq \int_{2m_\pi}^{\infty}dy\,d^{\text{Sill}}_{\rho}(y) \,\Gamma_{\rho_3\rightarrow \rho\eta}(m_{\rho_3},y,m_{\eta})\simeq (3.95\pm0.81) \,\text{MeV .}
\end{align}
By using the decay rates given in Tab.\ref{tab:rho3}, we present various three-body decay channels assuming the \text{Sill} distribution in Tab. \ref{tab:3body}.

\begin{table}[h] %[ptb]
		\centering
		\renewcommand{\arraystretch}{1.5}
		\begin{tabular}[c]{|c|c|c|}
			\hline
			Decay channel  & \text{Sill} distribution &PDG \\\hline
   $\,\;\pi_2(1670)\rightarrow\rho(770)\pi \rightarrow \pi^{\pm}\pi^{+}\pi^{-}$ &  $73.9\pm 9.9 $ & $137\pm 11  $  \\ \hline
    $\,\;\pi_2(1670)\rightarrow \overline{K}^{\star}(892)\pi \rightarrow K\overline{K}\pi $ & $10.4\pm 1.4$  & \\\hline
   $\,\;\rho_3(1690)\rightarrow \rho(770)\eta \rightarrow \eta\pi\pi$ &  $3.88\pm 0.80 $ & seen  \\ \hline
   $\,\;\rho_3(1690) \rightarrow \overline{K}^{\star}(892)\pi \rightarrow K\overline{K}\pi $ & $3.43\pm 0.70$  & $6.1\pm 2.0 $\\\hline
%  $K_{3}^{\ast} (1780) \rightarrow K\pi\pi $									&	$16.26\pm 3.34 $					&	$ $
		%\\\hline
		%$K_{3}^{\ast} (1780) \rightarrow K\pi \, \pi$						&	$25.54\pm 5.25 $					&	$ $\\\hline

   \end{tabular}
    \caption{Three body decay rates of $\pi_2(1670)$ and $\rho_3(1690)$ mesons (MeV).}
      \label{tab:3body}
   \end{table}

%   \section{Three body decays of $K_{3}^{\ast} (1780)$ }

 %  \section{Three body decays of $G_3$}

\section{Conclusion}
In this paper, we used a hadronic model and the Sill distribution, introduced in the vector meson intermediate decays, to evaluate the three-body decays of $\pi_2(1670)$ and $\rho_3(1690)$. Results are compared to the ones listed in PDG: they agree within the error bars for $\rho_3(1690)\rightarrow K\overline{K}\pi$, while it is two times off the experimental result for $\pi_2(1670)\rightarrow 3\pi$. This implies the necessity of the consideration of multi-channel version of the \text{Sill} distribution \cite{Giacosa:2021mbz} in the future.

\section*{Acknowledgement}
We are thankful to Francesco Giacosa, Adrian Koenigstein, and Vanamali Shastry for useful discussions. We acknowledge
financial support through the project “Development Accelerator of the Jan Kochanowski University of
Kielce”, co-financed by the European Union under the European Social Fund, with no. POWR.03.05.00-00-Z212 / 18. The work of Shahriyar Jafarzade is partially supported by the Polish National Science Centre (NCN) through the OPUS
project 2019/33/B/ST2/00613.

\bibliographystyle{utphys}
\bibliography{references}

\providecommand{\href}[2]{#2}\begingroup\raggedright\begin{thebibliography}{10}

\bibitem{Workman:2022ynf}
{\bfseries Particle Data Group} Collaboration, R.~L. Workman {\em et~al.}, ``{Review of Particle Physics},'' \href{http://dx.doi.org/10.1093/ptep/ptac097}{{\em PTEP} {\bfseries 2022} (2022) 083C01}.

\bibitem{Koenigstein:2016tjw}
A.~Koenigstein and F.~Giacosa, ``{Phenomenology of pseudotensor mesons and the pseudotensor glueball},'' \href{http://dx.doi.org/10.1140/epja/i2016-16356-x}{{\em Eur. Phys. J. A} {\bfseries 52} no.~12, (2016) 356}, \href{http://arxiv.org/abs/1608.08777}{{\ttfamily arXiv:1608.08777 [hep-ph]}}.

\bibitem{Giacosa:2023fdz}
F.~Giacosa, S.~Jafarzade, and R.~Pisarski, ``{Anomalous interactions for heterochiral mesons with $J^{PC}=1^{+-}$ and $2^{-+}$},'' \href{http://arxiv.org/abs/2309.00086}{{\ttfamily arXiv:2309.00086 [hep-ph]}}.

\bibitem{Jafarzade:2021vhh}
S.~Jafarzade, A.~Koenigstein, and F.~Giacosa, ``{Phenomenology of $J^{PC}$ = $3^{--}$ tensor mesons},'' \href{http://dx.doi.org/10.1103/PhysRevD.103.096027}{{\em Phys. Rev. D} {\bfseries 103} no.~9, (2021) 096027}, \href{http://arxiv.org/abs/2101.03195}{{\ttfamily arXiv:2101.03195 [hep-ph]}}.

\bibitem{Shastry:2021asu}
V.~Shastry, E.~Trotti, and F.~Giacosa, ``{Constraints imposed by the partial wave amplitudes on the decays of J=1, 2 mesons},'' \href{http://dx.doi.org/10.1103/PhysRevD.105.054022}{{\em Phys. Rev. D} {\bfseries 105} no.~5, (2022) 054022}, \href{http://arxiv.org/abs/2107.13501}{{\ttfamily arXiv:2107.13501 [hep-ph]}}.

\bibitem{Parganlija:2012fy}
D.~Parganlija, P.~Kovacs, G.~Wolf, F.~Giacosa, and D.~H. Rischke, ``{Meson vacuum phenomenology in a three-flavor linear sigma model with (axial-)vector mesons},'' \href{http://dx.doi.org/10.1103/PhysRevD.87.014011}{{\em Phys. Rev. D} {\bfseries 87} no.~1, (2013) 014011}, \href{http://arxiv.org/abs/1208.0585}{{\ttfamily arXiv:1208.0585 [hep-ph]}}.

\bibitem{Giacosa:2021mbz}
F.~Giacosa, A.~Okopi\'nska, and V.~Shastry, ``{A simple alternative to the relativistic Breit\textendash{}Wigner distribution},'' \href{http://dx.doi.org/10.1140/epja/s10050-021-00641-2}{{\em Eur. Phys. J. A} {\bfseries 57} no.~12, (2021) 336}, \href{http://arxiv.org/abs/2106.03749}{{\ttfamily arXiv:2106.03749 [hep-ph]}}.

\bibitem{Jafarzade:2022uqo}
S.~Jafarzade, A.~Vereijken, M.~Piotrowska, and F.~Giacosa, ``{From well-known tensor mesons to yet unknown axial-tensor mesons},'' \href{http://dx.doi.org/10.1103/PhysRevD.106.036008}{{\em Phys. Rev. D} {\bfseries 106} no.~3, (2022) 036008}, \href{http://arxiv.org/abs/2203.16585}{{\ttfamily arXiv:2203.16585 [hep-ph]}}.

\bibitem{Shastry:2023ths}
V.~Shastry and F.~Giacosa, ``{Radiative production and decays of the exotic \ensuremath{\eta}1'(1855) and its siblings},'' \href{http://dx.doi.org/10.1016/j.nuclphysa.2023.122683}{{\em Nucl. Phys. A} {\bfseries 1037} (2023) 122683}, \href{http://arxiv.org/abs/2302.07687}{{\ttfamily arXiv:2302.07687 [hep-ph]}}.

\bibitem{Winney:2022tky}
{\bfseries Joint Physics Analysis Center} Collaboration, D.~Winney, A.~Pilloni, V.~Mathieu, A.~N. Hiller~Blin, M.~Albaladejo, W.~A. Smith, and A.~Szczepaniak, ``{XYZ spectroscopy at electron-hadron facilities. II. Semi-inclusive processes with pion exchange},'' \href{http://dx.doi.org/10.1103/PhysRevD.106.094009}{{\em Phys. Rev. D} {\bfseries 106} no.~9, (2022) 094009}, \href{http://arxiv.org/abs/2209.05882}{{\ttfamily arXiv:2209.05882 [hep-ph]}}.

\bibitem{Yildirim:2023znd}
D.~Y\i{}ld\i{}r\i{}m, ``{Spin partners of the $B^{(*)}\bar{B}^{(*)}$ resonances with a different approach than the Breit\textendash{}Wigner parameterization},'' \href{http://dx.doi.org/10.1140/epja/s10050-023-01061-0}{{\em Eur. Phys. J. A} {\bfseries 59} no.~7, (2023) 148}, \href{http://arxiv.org/abs/2306.12403}{{\ttfamily arXiv:2306.12403 [hep-ph]}}.

\bibitem{ALICE:2023wjz}
{\bfseries ALICE} Collaboration, S.~Acharya {\em et~al.}, ``{Accessing the strong interaction between \ensuremath{\Lambda} baryons and charged kaons with the femtoscopy technique at the LHC},'' \href{http://dx.doi.org/10.1016/j.physletb.2023.138145}{{\em Phys. Lett. B} {\bfseries 845} (2023) 138145}, \href{http://arxiv.org/abs/2305.19093}{{\ttfamily arXiv:2305.19093 [nucl-ex]}}.

\bibitem{Giacosa:2023xti}
F.~Giacosa and V.~Shastry, ``{Sill distribution: genesis and salient features},'' in {\em {20th International Conference on Hadron Spectroscopy and Structure}}.
\newblock 10, 2023.
\newblock \href{http://arxiv.org/abs/2310.06346}{{\ttfamily arXiv:2310.06346 [hep-ph]}}.

\bibitem{Fukui:1988mp}
S.~Fukui {\em et~al.}, ``{Vector Resonances Around 1.6-{GeV} of the $\eta \pi^+ \pi^-$ System in the $\pi^- p$ Charge Exchange Reaction at 8.95-{GeV}/$c$},'' \href{http://dx.doi.org/10.1016/0370-2693(88)90500-X}{{\em Phys. Lett. B} {\bfseries 202} (1988) 441--446}.

\end{thebibliography}\endgroup

\end{document}